\documentclass[11pt,twoside]{article}

\usepackage{asp2006}
\usepackage{epsf}
\usepackage{epsfig}
\usepackage{lscape}

\begin{document}
\title{Properties of the Narrow-Line Region in Seyfert Galaxies}   
\author{Nicola Bennert}  
\affil{Institute of Geophysics and 
Planetary Physics, University of California, Riverside, CA 92521, USA;
nicola.bennert@ucr.edu}    
\author{Bruno Jungwiert}   
\affil{Astronomical Institute, Academy of Sciences of the Czech Republic,
Bo{\v c}n\'\i\ II 1401, 141 31 Prague 4, Czech Republic}  
\author{Stefanie Komossa}  
\affil{Max-Planck Institut f\"ur extraterrestrische Physik,
              Giessenbachstrasse 1, D-85748 Garching, Germany} 
\author{Martin Haas, Rolf Chini}   
\affil{Astronomisches Institut Ruhr-Universit\"at Bochum,
  Universit\"atsstrasse 150, D-44708 Bochum, Germany}

\begin{abstract} 
We study the narrow-line region (NLR)
of six Seyfert-1 and six Seyfert-2 galaxies
by means of spatially resolved
optical spectroscopy and photoionization modelling.
From spatially resolved spectral diagnostics, we find a 
transition between the AGN-excited NLR and the surrounding 
star-forming regions, 
allowing us to determine the NLR size independent of stellar contamination. 
CLOUDY photoionization models show that the observed transition 
represents a true difference in ionization source and 
cannot be explained by variations 
of physical parameters. The electron density and 
ionization parameter decrease with radius 
indicating that the NLR is photoionized by the central source only. 
The velocity field suggests a disky NLR gas distribution.
\end{abstract}

\section*{The Narrow-Line Region in Active Galaxies}   
The narrow-line region (NLR) in active galactic nuclei (AGNs)
is ideally suited to study the central region 
through its various emission lines. It has the great advantage over the 
broad-line region (BLR) to be spatially resolvable, at least for nearby AGNs.
Since the NLR is affected by various parameters such as the energy input from the 
central engine, the existence of a nuclear torus, radio jets, and star 
formation, it accesses a number of key questions of AGN physics. 
 
[\ion{O}{iii}]\,$\lambda$5007\,\AA~(hereafter 
[\ion{O}{iii}]) narrow-band imaging is commonly used to study the NLRs of 
active galaxies, especially Seyferts (e.g. Mulchaey et al. 1996a, Falcke et al. 1998). 
For QSOs, this method was first applied by
\citet{ben02} using an HST emission-line imaging survey
of seven radio-quiet PG QSOs with z$<$0.5.
Including seven Seyferts, they discovered a relation between NLR size and [\ion{O}{iii}]
luminosity.
This newly found NLR size-luminosity relation connects the NLR of QSOs and Seyferts 
over three orders of magnitude. 
This relation seems to continue the relation found between BLR size
and luminosity [e.g.~\citet{kas00}], indicating that both BLR and 
NLR size are only determined by the AGN luminosity.
Indeed both sizes are related, 
providing the potential to estimate black hole (BH) masses from the 
NLR size \citep{ben04,ben05}. Still, the question remains whether
the NLR size-luminosity relation holds at high luminosity or whether
these objects lose their very large, dynamically unbound NLR \citep{net04}.

However, the slope of the NLR size-luminosity relation has been controversial. 
While \citet{ben02} obtained a slope of ~0.5
suggesting a self-regulating mechanism that determines the size to scale with
the ionization parameter, \citet{sch03} report a 
slope of 0.33 for their sample of 60 Seyfert galaxies imaged with HST. Such a slope is
expected for a homogeneous gaseous sphere
ionized by a central source (Str\"omgren sphere).
When comparing the two samples and calculating a fit to all type-1 and type-2 AGNs 
separately, the relation for type-2 AGNs is close to 0.33, 
while type-1 objects are fit with a slope of 0.55 \citep{ben05, ben06c}. 
This result explains the different slopes found by the two groups:
In the sample of \citet{ben02}, type-1 QSOs determine the slope whereas the 
slope of the sample studied by \citet{sch03} is dominated by Sy2s. 
The difference can be explained by the observers viewing angle in the 
framework of the unified model and a receding torus \citep{ben05, ben06c}.

One problem that remains when studying the NLR by 
[\ion{O}{iii}] imaging alone is that this emission can be contaminated by 
contributions from e.g.~star formation or shock-ionized gas, altering the NLR size. 
In addition, the measured size depends on the depth of the images:
When comparing
ground-based [\ion{O}{iii}] images of Seyfert galaxies
from \citet{mul96} with the HST snapshot survey of \citet{sch03},
the latter reveal, on average, six times smaller NLR sizes, probably due to 
the 15 to 20 times lower sensitivity.
These considerations question the determination of the ``NLR size'' from 
[\ion{O}{iii}] imaging alone.

\section*{Answers from Spatially Resolved Spectral Diagnostics}   
Spatially resolved spectroscopy
is a valuable alternative approach as it can directly probe
the size in terms of AGN photoionization and discriminate
the stellar or shock-ionized contribution by making use of selected emission-line ratios.
In addition, several physical parameters of the NLR such as ionization parameter,
electron density, and velocity can be directly
accessed and analysed as a function of distance from the nucleus.

Therefore, we studied the NLR of six Sy1 and six Sy2 galaxies by means of 
optical longslit spectroscopy with VLT/FORS1 and NTT/EMMI.
In the following, we summarize these results, focussing on the nearby
Seyfert-2 galaxy NGC\,1386 which can be considered as prototypical
for the whole sample. For a detailed discussion of the
sample properties, observations, data reduction and analysis, as well as results,
see Bennert et al. (2005, 2006a,b,d).

We use the galaxy itself,
to subtract the underlying stellar absorption lines,
applying reddening corrections for the
increasing amount of dust towards the center.

\begin{figure}
 \resizebox{\hsize}{!}{\includegraphics[angle=-90]{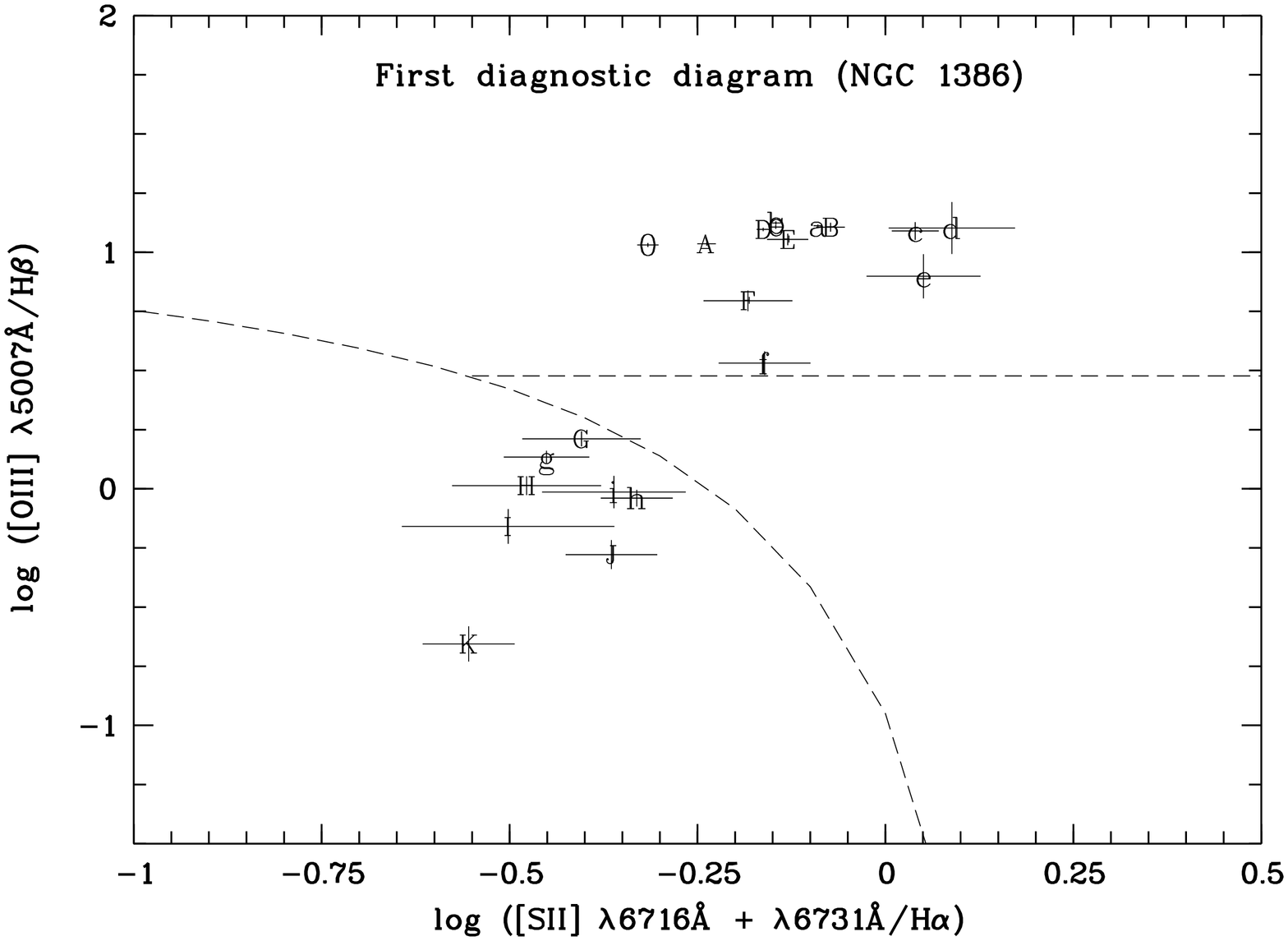}
\includegraphics[angle=-90]{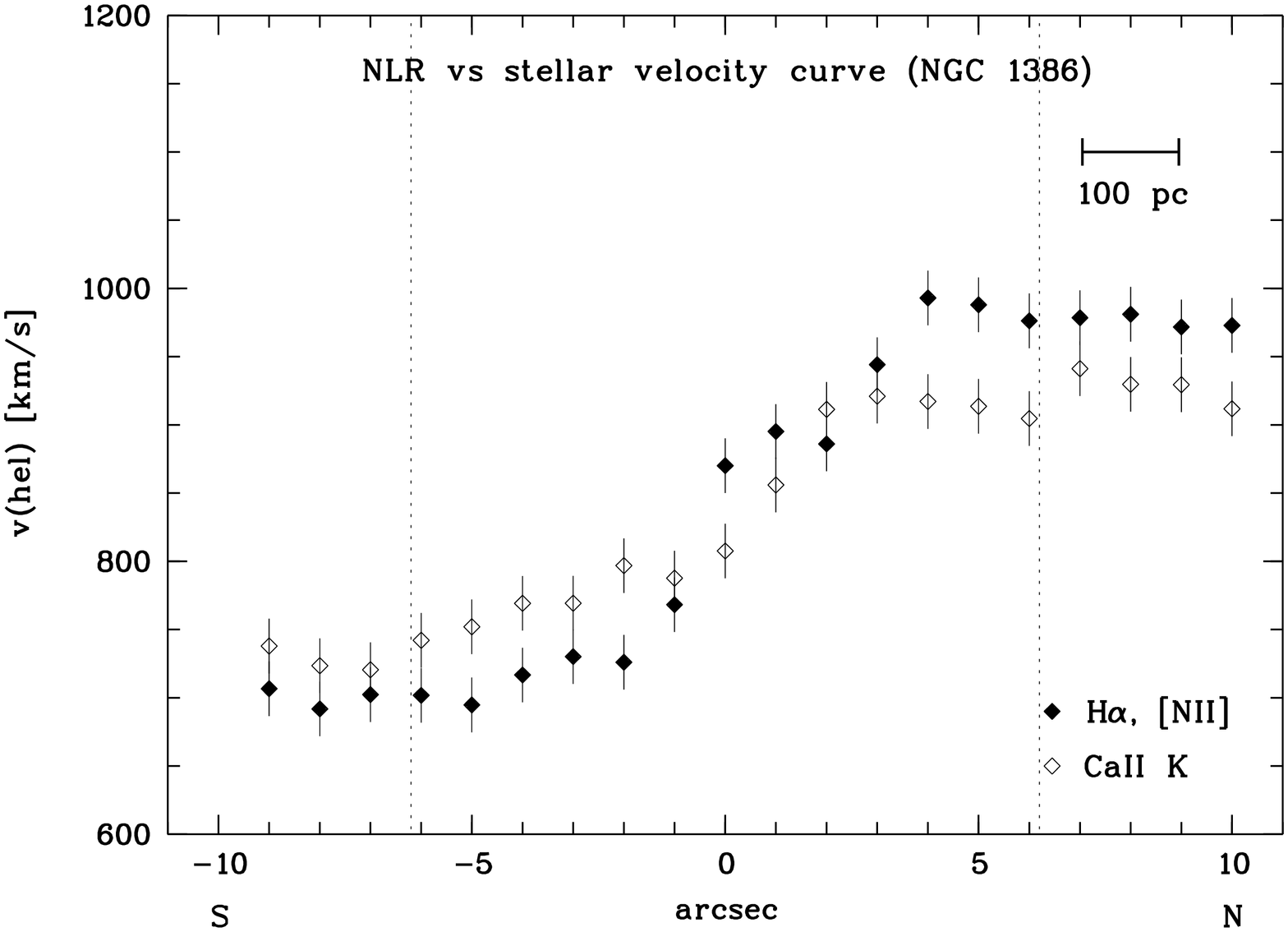}}
\caption{\small{\label{diag} {\it Left panel:} Diagnostic diagram for
spatially resolved line ratios.
The dividing lines were taken from the analytic AGN diagnostics
of \citet{kew01}. The symbols are chosen such that ``0'' refers to
the central spectrum, the small letters mark southern regions, the capital
ones northern regions (``a,A'' = 1 arcsec distance from the nucleus = 52\,pc, 
``b,B'' = 2\arcsec~distance = 104\,pc etc.). 
Only the central $r \sim$ 6\arcsec~show 
ratios expected for AGN-photoionized gas (upper right corner).
Further out, the line ratios fall
into the \ion{H}{ii}-region regime (lower left corner). {\it Right panel:} 
Velocity curve of
the NLR derived from the average value of the peak wavelengths
of the H$\alpha$ and [\ion{N}{ii}] emission lines (filled symbols) 
as well as the stellar 
velocity curve measured using the \ion{Ca}{ii} K absorption-line 
``peak'' wavelength as seen in the ``raw'' spectrum (open symbols).
 The edge of the NLR as determined from the
  diagnostic diagrams is indicated by dotted lines.}
}
\end{figure}

From spatially resolved spectral diagnostics,
we find a transition between central line ratios falling into the
AGN regime and outer ones in the \ion{H}{ii}-region regime for
four objects in our sample (two Sy1s, two Sy2s; Fig.~\ref{diag}
({\it left panel}).
Applying \texttt{CLOUDY} \citep{fer98} photoionization models, we show that the observed distinction
between \ion{H}{ii}-like and AGN-like ratios in NGC\,1386 represents a true
difference in ionization source and cannot
be explained by variations of physical parameters such as
ionization parameter, electron density or metallicity \citep{ben06a}. We 
interpret it as a real border between the NLR,
i.e. the central AGN-photoionized region, and 
surrounding \ion{H}{ii} regions.
It allows us to determine the NLR size free of stellar contamination,
yielding an NLR radius of 6\arcsec. This is
twice as large as the one measured from the \citet{sch03} [\ion{O}{iii}] HST image,
due to the low sensitivity of their snap-shot survey.
On the other hand, it is half the value derived by \citet{fra03}
from long-slit spectroscopy as their NLR size 
includes [\ion{O}{iii}] emission
from surrounding \ion{H}{ii} regions.

\begin{figure}
\centering
 \resizebox{\hsize}{!}{\includegraphics[angle=-90]{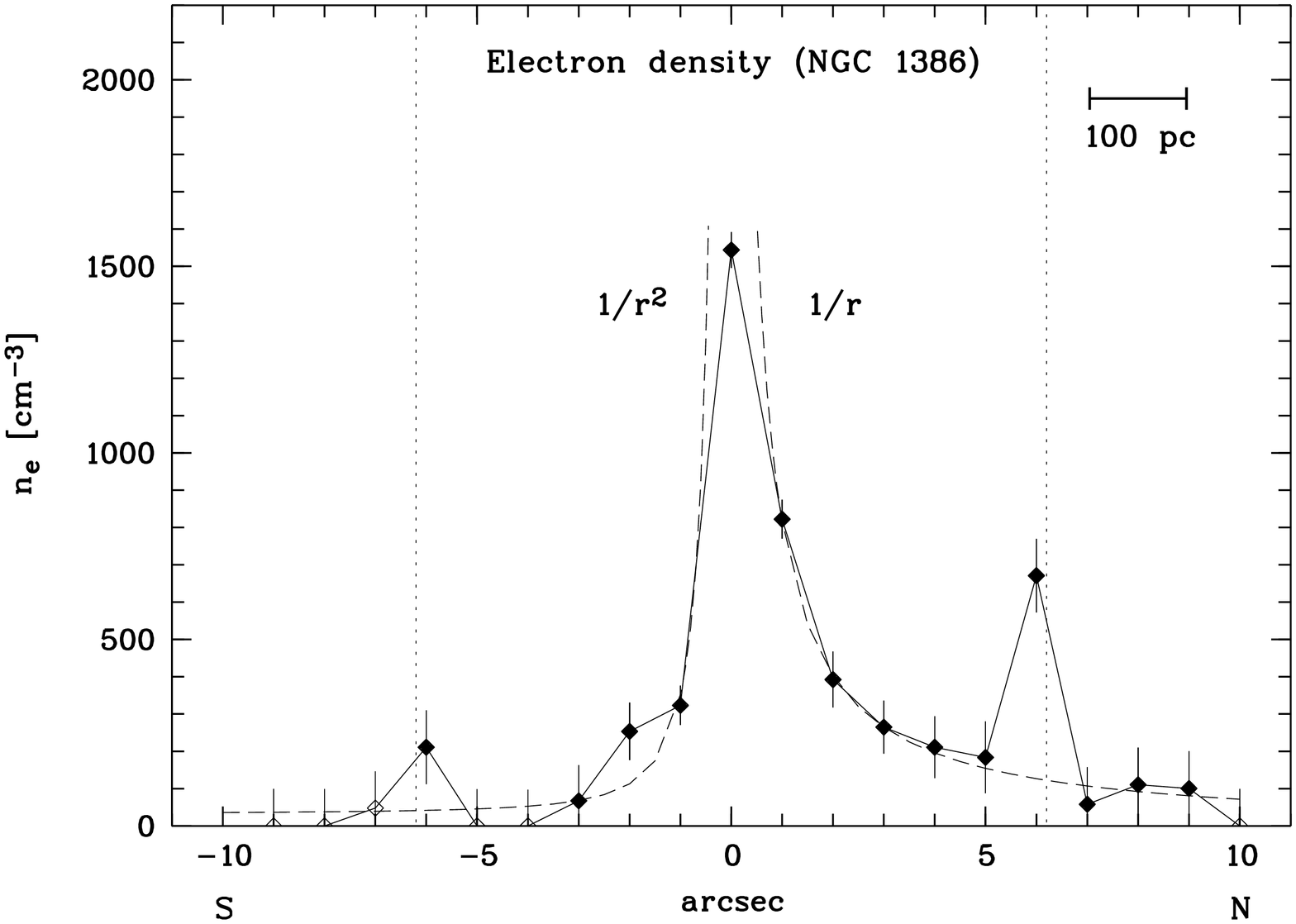}
\includegraphics[angle=-90]{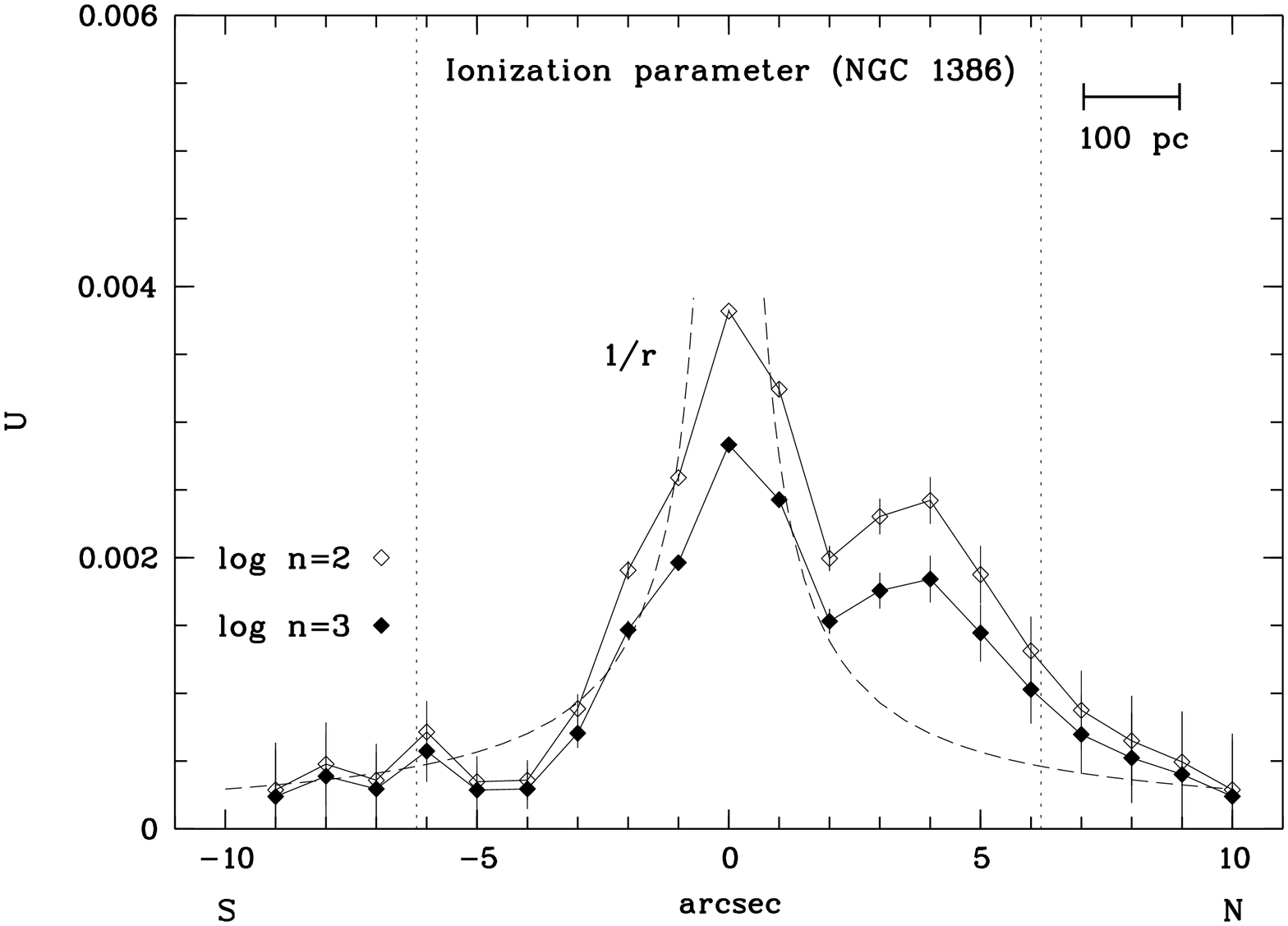}}
\caption{{\small \label{density} {\it Left panel:} Electron density obtained
from the [SII]\,$\lambda$6716\,\AA/$\lambda$6731\,\AA~ratio as a function of distance from the nucleus.
Open symbols indicate locations
where $n_e$ is in the low-density limit (assumed $\le$50\,cm$^{-3}$).
For comparison, the powerlaws $n_e (r) \propto n_{e,0}
  r^{-1}$ and $n_e (r) \propto n_{e,0}
  r^{-2}$ are shown as dashed lines. 
{\it Right panel:} Ionization parameter
derived from [OII]\,$\lambda$3727\,\AA/[\ion{O}{iii}]\,$\lambda$5007\,\AA~ratio
as a function of distance from the nucleus (open symbols: $n_H$ =
100\,cm$^{-3}$, filled ones: $n_H$ = 1000\,cm$^{-3}$).
A powerlaw $U(r) \propto U_{0}
  r^{-1}$ is shown as dashed line. In both panels, the edge of the NLR as determined from the
  diagnostic diagrams is indicated by dotted lines.}
}
\end{figure}

For almost all 12 Seyfert galaxies,
we find that both the electron density and 
the ionization parameter decrease with
roughly $r^{-1}$ (Fig.~\ref{density}). 
This is in agreement with photoionization of the NLR by 
a central source, 
supporting our approach to determine the NLR size via diagnostic diagrams.

In most objects, 
the NLR and stellar velocity fields are similar and indicative
of rotation (Fig.~\ref{diag}, {\it right panel}), implying that the NLR
gas is distributed in a disk.

\acknowledgements 
N.B. and S.K. thank the organizers for invitation and
hospitality. N.B. acknowledges 
the support of the American Astronomical Society 
and the National Science Foundation in the form 
of an International Travel Grant. 
B.J. acknowledges the support 
of the Czech Research Project No. AV0Z10030501.
M.H. is supported by ``Nordrhein-Westf\"alische
Akademie der Wissenschaften''.

\end{document}